\documentclass[a4paper,aps,prr,showpacs,twocolumn]{revtex4-1}
\usepackage{amsmath,amssymb,mathrsfs,dsfont,color,hyperref}
\usepackage[pdftex]{graphicx}
\usepackage[dvipsnames]{xcolor}
\usepackage{subfigure}
\usepackage[normalem]{ulem}

\newcommand\ContFracOp{%
  \operatornamewithlimits{%
    \mathchoice
     {\vcenter{\hbox{\huge $\mathcal{K}$}}}
     {\vcenter{\hbox{\Large $\mathcal{K}$}}}
     {\mathcal{K}}
     {\mathcal{K}}}}

\begin{document}
\bibliographystyle{unsrt}

\title{Thermal topological phase transition in SnTe from \emph{ab-initio} calculations}

\author{Pablo Aguado-Puente}
\affiliation{CIC Nanogune BRTA, E-20018 Donostia - SanSebastian, Spain}

\author{Piotr Chudzinski}
\affiliation{School of Mathematics and Physics, Queen's University of Belfast, Belfast, Northern Ireland (United Kingdom)}
\affiliation{Institute of Fundamental Technological Research, Polish Academy of Sciences, Pawinskiego 5B, PL-02-105 Warsaw, Poland}

\date{22.02.2022}

\pacs{}

\begin{abstract}

	One of the key issues in the physics of topological insulators is whether the topologically non-trivial properties survive at finite temperatures and, if so, whether they disappear only at the temperature of topological gap closing. Here, we study this problem, using quantum fidelity as a measure, by means of \emph{ab-initio} methods supplemented by an effective dissipative theory built on the top of the \emph{ab-initio} electron and phonon band structures.  In the case of SnTe, the prototypical crystal topological insulator, we reveal the presence of a characteristic temperature, much lower than the gap-closing one, that marks a loss of coherence of the topological state. The transition is not present in a purely electronic system but it appears once we invoke coupling with a dissipative bosonic bath. Features in the dependence with temperature of the fidelity susceptibility can be related to changes in the band curvature, but signatures of a topological phase transition appear in the fidelity only though the non-adiabatic coupling with soft phonons. Our argument is valid for valley topological insulators, but in principle can be generalized to the broader class of topological insulators which host any symmetry-breaking boson.   

\end{abstract}

\maketitle

{\it Introduction.} The theoretical description \cite{Kane2005} and subsequent observation \cite{Konig2007} of topological states in the early 2000s have dramatically changed the landscape of research in condensed matter physics. Thousands of theoretical papers have been published on this subject, with an overwhelming majority dedicated to the properties of zero-temperature model systems \cite{Qi2011}. The question of finite-temperature effects is a very difficult one and has been put in the spotlight only a few years ago \cite{Viyuela2014,Viyuela2015}.
Naturally, the issue of the theoretical treatment of thermal effects in topological insulators is a pressing one since all the experiments are carried out at finite temperature. However, the standard Ginzburg-Landau theory of phase transitions is inapplicable here -- one cannot define a local order parameter, instead one must work with invariants defined along entire trajectories in the material's parameter space \cite{Kempkes2016Hill}. 
There have been so far several theoretical attempts to tackle this problem. From a single-particle density functional theory (DFT) perspective, attempts to assess the topological critical temperature are generally based on the closing of the gap at some finite temperature \cite{Monserrat2016,Antonius2016,Querales-Flores2020}. The conjuncture made in these papers is that topological invariants stay unaltered up to this point and the entire physics is determined by the single-particle gap inversion. 
Since the topological invariants result from the collective behavior of the entire electronic liquid, this assumption needs to be taken with care.
The picture provided by many-body methods is in fact strikingly different. First attempts used a concept of Uhlmann parallel transport which generalized a concept of a finite-temperature Chern number and aimed to compute its temperature dependence \cite{Viyuela2014}. 
These studies were initially limited to quantum one-dimensional (1D) systems \cite{Viyuela2014,Andersson2016} and later generalized to 2D \cite{Viyuela2015}. 
This kind of analysis, applied to model systems, revealed that the critical temperature can be up to 70\% lower than the mean-field critical temperature \cite{Viyuela2014}. 
There have also been attempts to apply Hill's thermodynamics (small quantum system thermodynamics) to capture the influence of the thermodynamic potential of the boundary terms \cite{Quelle2016Hill, Kempkes2016Hill}. These numerical studies confirmed the results of the Uhlmann conjecture in model 2D systems and extended the observations to the 3D case.  
Recently, the use of fidelity was proposed \cite{Mera2017,VieiraAmin} as a suitable tool to probe the quantum phase transitions even in systems with non-local order parameters. An analysis based on this magnitude suggests that in a topological model the fidelity only shows signs of a phase transition at zero temperature \cite{Mera2017,VieiraAmin}. This result would imply that topological states are a characteristic of zero-temperature states and disappear in a crossover manner as the temperature increases, revealing a lack of protection against thermal fluctuations of the topological state of matter. However, in Ref. \cite{Mera2017} only Hamiltonians with temperature-independent parameters were considered. 

Here, we address this issue by computing the fidelity on a realistic system with a temperature-dependent electronic structure. We consider a topological insulator protected by the valley (mirror) symmetry of the underlying crystal lattice \cite{Fu2011} as we expect that in this case the contributions from thermal fluctuations of the lattice will manifest particularly strongly. We study SnTe, which is perhaps the simplest compound of this kind \cite{Hsieh2012}. Being material specific allows us to perform a joint study that builds an extension of \emph{ab-initio} results by an exact analytic calculation of drag effects.

\begin{figure}[h!]
    \centering
    \includegraphics[width=0.99\columnwidth]{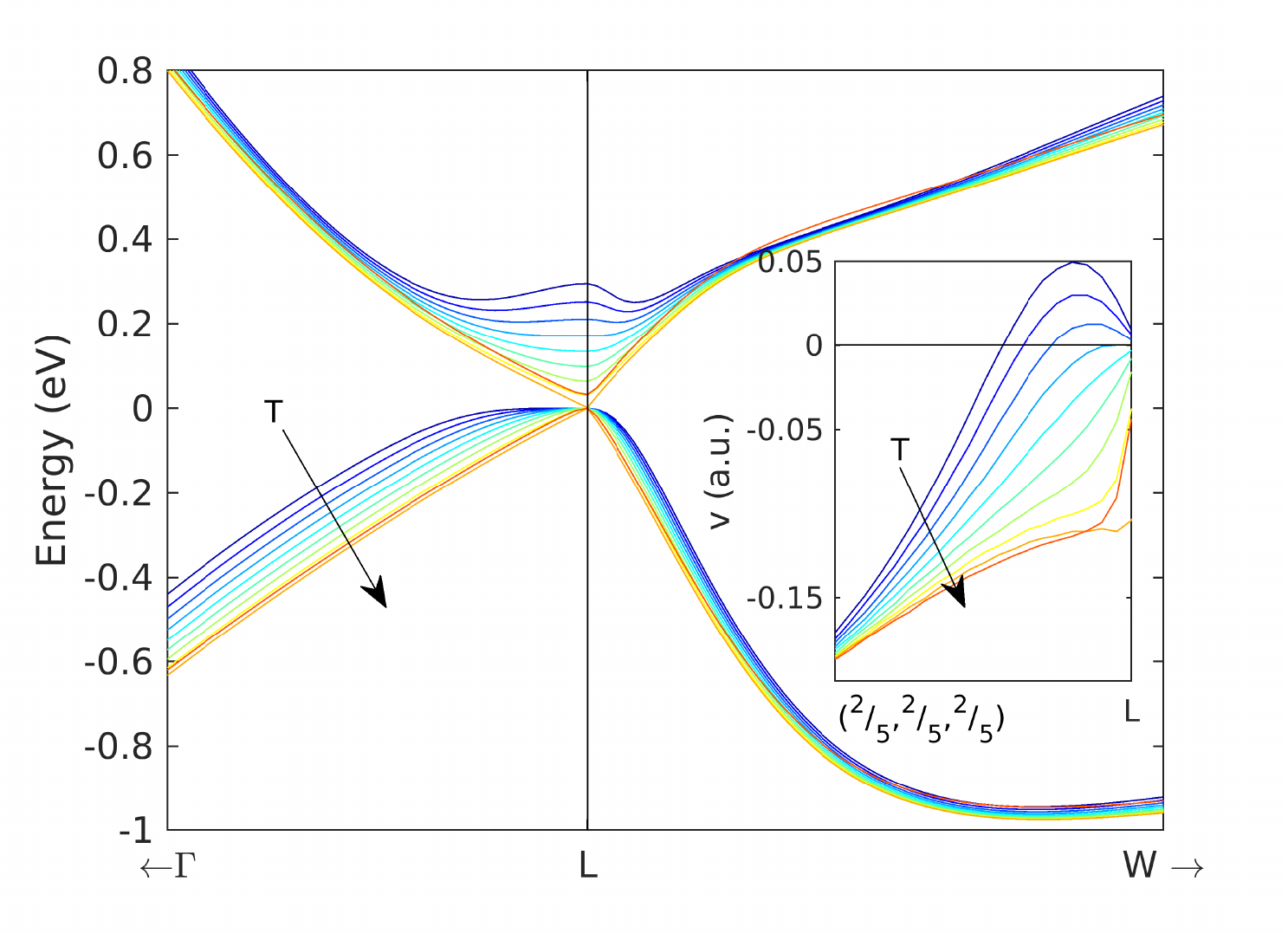}
    \caption{ Detail of the band structure around the $L$ point showing the evolution of the band structure of SnTe with temperature, as calculated with DFT (the band structure along a full path across the Brillouin zone can be found in Appendix \ref{sec:app_DFT}). The temperature, $T$ varies from 100 to 1900 K in 200-K steps. Inset: Group velocities of electrons in the conduction band as a function of temperature.}
    \label{fig:bands}
\end{figure}

{\it Band structure as provided by DFT}. Our starting point is the density functional theory (DFT) band structure computed using the PBE parametrization for the exchange and correlation functional Technical details of the DFT simulations can be found in Appendix \ref{sec:app_DFT}. The choice of the Perdew-Burke-Ernzerhof (PBE) functional over higher-order methods, such as hybrid functionals or $GW$, is justified by previous studies, which show that the effect of additional electronic correlation and band shift due to an electron-phonon interaction tend to partially cancel each other \cite{Aguado-Puente2020,Querales-Flores2020}.
The explicit changes of the electronic structure due to thermal expansion are taken into account by performing calculations at the temperature-dependent lattice parameter, obtained from the linear thermal expansion coefficient as calculated in Ref. \cite{Querales-Flores2020}.
The electronic dispersion curves near the band edges at $L$ are displayed in Fig. \ref{fig:bands} as a function of temperature. At low temperatures the bottom of the conduction is shifted with respect to the $L$ point, showing a ``Mexican-hat''-like feature that, in SnTe, reflects the topological character that results from the inversion of bands of opposite parity close to the $L$ point \cite{Ye2015}. As the temperature increases, and the material expands, the gap tends to close. In the process, the ``Mexican-hat''-like minimum disappears and the dispersion becomes linear near the crossover. Neglecting the effect of electron-phonon coupling, according to the DFT simulations the closing of the gap would occur at a temperature of around 1700 K due to lattice expansion exclusively, a temperature larger than the melting temperature of the material (1063 K). 

{\it Computing fidelity}. 
We follow the method introduced in Ref. \cite{Mera2017} to track the topological phase transition and use the fidelity, defined as
\begin{equation}
  \mathcal{F}(\rho^A, \rho^B) = Tr\left[\sqrt{\sqrt{\rho^A}\rho^B\sqrt{\rho^A}}\right],
  \label{eq:fidelitydef}
\end{equation}
where $\rho^I$ are the corresponding density matrices. 
The fidelity is a generalization of the overlap for mixed states and measures the similarity between two quantum states \cite{Steane1998}. At a phase transition the fidelity drops as a result of the drastic changes in the quantum state of the system \cite{Zanardi2006}. Here we use the fidelity to measure the proximity between states of the system at $T$ and $T+\delta T$ in order to detect a hypothetical phase transition at finite temperature.

We first compute the single-particle fidelity as a function of temperature using the density matrices obtained from DFT calculations. The single-particle density matrix is readily available in many DFT codes, where it is commonly expressed in the basis of the basis functions used to expand the Kohn-Sham states. In this basis the density matrix takes the form
\begin{equation}
    \rho = \sum_{\nu,\mu} \rho_{\nu,\mu} |\nu\rangle\langle\mu|,
\end{equation}
with
\begin{equation}
    \rho_{\nu,\mu} = \sum_{n\mathbf{k}} f_{n\mathbf{k}} c_{\nu,n\mathbf{k}} c_{\mu,n\mathbf{k}}^*,
\end{equation}
where $\{|\nu\rangle, |\mu\rangle \}$ are basis orbitals, $n\mathbf{k}$ denotes a Kohn-Sham state, and $f_{n\mathbf{k}}$ is the occupation function. 

In the analysis of the topological character of the electronic structure based on zero-temperature topological invariants, temperature effects are usually limited to lattice expansion and band renormalization due to electron-phonon effects \cite{Monserrat2016,Antonius2016,Querales-Flores2020}. Instead, the analysis based on the quantum fidelity (density matrices) also accounts for changes in the topological character due to the partial occupation of eigenvalues of opposite parity. Remarkably, despite the notable changes in the band dispersion of the material in this temperature range, the single-particle fidelity is not able to capture any sign of the transition, even at the crossover temperature ($\sim 1700$ K). 

\begin{figure}[hb]
    \centering
     \includegraphics[width=0.9\columnwidth]{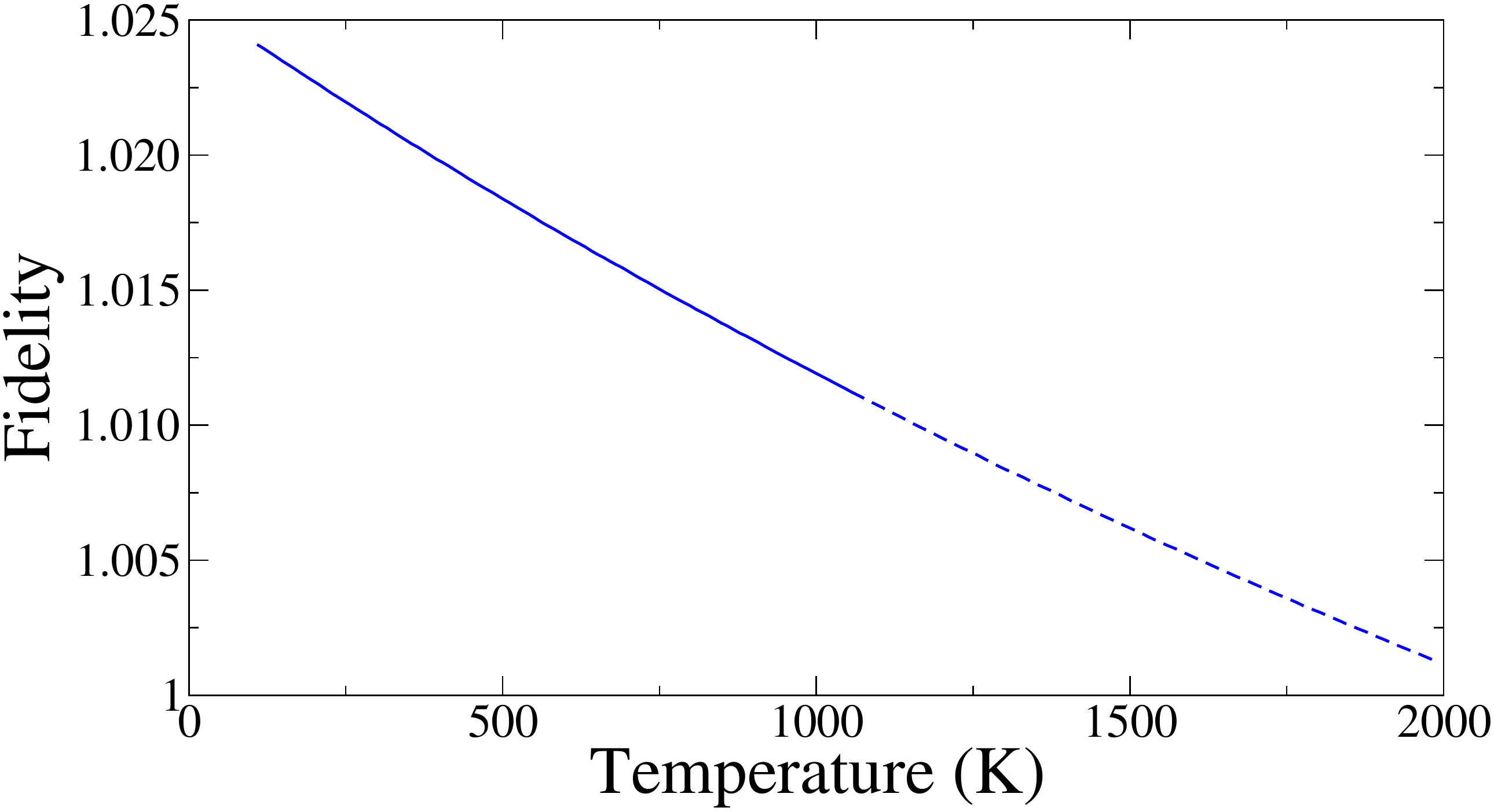}
    \caption{Fidelity as a function of temperature computed from the single-particle DFT density matrix. Data for temperatures above the melting point, and therefore not experimentally accessible, are plotted as a dashed line. }
    \label{fig:fidelity}
\end{figure}

{\it Dissipative environment}. In SnTe the crystal lattice determines the mirror symmetry by which the topological invariant is defined. At low temperature the material undergoes a ferroelectric phase transition, and electron-phonon coupling is expected to be substantial (as is the case in similar materials \cite{Jiang2016}), so it is therefore necessary to introduce the lattice dynamics explicitly into our considerations. The motion of atoms associated with a soft transverse optical (TO) mode at $\Gamma$ violates the mirror symmetry with respect to which the topological invariant is defined. If dissipation is introduced by a dynamic electron-phonon coupling, then part of the electronic density is brought into an incoherent fluctuating state and the topological state is undefined. Here, we develop a model for an electron gas with a realistic dispersion given by the DFT bands in Fig. \ref{fig:bands} interacting nonadiabatically with a phonon bath. Such dissipative systems are known to host a phase transition \cite{Soriente-dissiPT,Quan-fidel-thermal}. The Hamiltonian for the electron plus phonon system is
\begin{equation}\label{eq:hamilt}
    H=H_{DFT}[c_k^{\dag}c_k] + \sum_{k,q} g(q)c_k^{\dag}c_{k-q}(a_q^\dag + a_{-q}) + \sum_q \omega_q a_q^\dag a_q,
\end{equation}
where $\omega_q=c_{ph}q+\omega_0$ is the TO phonon dispersion (a linear dispersion is a valid approximation over a relatively wide range of $q$ around $\Gamma$ for the TO mode in SnTe \cite{ONeill2017}), and $g(q)$ is the electron-phonon coupling. We take $g(q)\sim q^{\alpha}$, a generalization of a displacement potential. In SnTe, for the TO phonon mode $\alpha<1$  as evidenced by a DFT study \cite{Querales-Flores2019}. The ladder vertex correction can further renormalize the value of $\alpha$ \cite{KREIMER2006rainbow}.

A solution of Eq. (\ref{eq:hamilt}) at a given temperature $T_0$ would be a specific state $|\rho_{tot}\rangle=|\rho_{el},n_{ph}\rangle$. Since the TO phonons are coupled with other branches (which in turn also interact with lattice dislocations, interfaces, etc.), this is a dissipative system. We refrain from solving this nontrivial  problem \cite{Dutta-weird-stdstate}, as we are interested only in the \emph{change} of the incoherent part of electronic density as the temperature changes \cite{Soriente-dissiPT,VieiraAmin}. The form of this Hamiltonian will hold as long as there is no structural phase transition, which for SnTe above 100 K is justified.

The time derivative of a density matrix is \cite{Weimer-EPL}
\begin{equation}
    \frac{d\hat{\rho}_{el}}{dt} =\imath[\hat{H},\hat{\rho}_{el}]+\mathcal{R}[\hat{\rho}_{el}],
    \label{eq:density_derivative}
\end{equation}
a general form that encapsulates the Lindbladian superoperator \cite{Manzano-Lindblad-rev} for a specific choice of a relaxation functional $\mathcal{R}[\hat{\rho}]$. It has been shown \cite{Weimer-var-princ} that, in general, for any steady state the norm of the left-hand side (LHS) of Eq. (\ref{eq:density_derivative}) (e.g. $|\ldots|=Tr\hat{\rho}_{el}$) is minimized, which implies that currents flowing in the steady state are minimized. Even without detailed knowledge of $\hat{\rho}_{el}$, this condition suffices to state that the variation around the steady state, $\delta \rho_{el}$, of the LHS must be equal to zero. By noting that only the second term in the Hamiltonian in Eq. (\ref{eq:hamilt}) does not commute with $\hat{\rho}_{el}$ and dividing the relaxation into unitary adiabatic time evolution and dissipative non-adiabatic components we can write the following phenomenological master equation for the \emph{variation} of the density with temperature: 
\begin{multline}\label{eq:master}
     \langle \delta (\hat{F}_{abs}-\hat{F}_{emit})\rangle|_{k,q}-\\
     q\left(V_F(k)-c_{ph}\right)\delta \rho_{inc}(k,q;T)-\mathds{1}q \delta \rho_{coh}(k,q;T)=0.
\end{multline}

Here, we have assumed that the $q$th phonon has been involved in the process, through absorption/emission operators $\hat{F}_\mathrm{abs,emit}$, $V_F$ is the Fermi velocity and $\rho_\mathrm{coh}$ and $\rho_\mathrm{inc}$ are the coherent and incoherent part of the density matrix. We have assumed a stationary condition $d\rho/dt=0$ and the two terms in the second line of Eq. (\ref{eq:master}) compensate the modified thermal ``drag'' force. This equation can be interpreted as an action of the Lindbladian superoperator on the density matrix $\hat{L}\{\delta \rho\}=0$. 
The first term in the Lindbladian is a scattering a term $\hat{F}_{i}\sim [H,\rho_\mathrm{tot}]$ and, according to Eq. (\ref{eq:hamilt}), electron liquid can constantly absorb and emit phonons \cite{Latour-phonon-duality}. We first consider a case with a \emph{single} phononic event when the emission-absorption drag ``force'' $\hat{F}$ is given by an analog of the Fermi golden rule,
\begin{equation}\label{eq:displacement}
    \langle \hat{F}_\mathrm{abs}-\hat{F}_\mathrm{emit}\rangle|_{k,q} = g_{el-ph}(q) N_q(f_{k-q}(1-f_{k})-f_k(1-f_{k-q})),
\end{equation}
where we took mean-field averages, i.e., $N_q$ and $f_k$ are the Bose-Einstein and Fermi-Dirac distribution functions, respectively. Increasing the temperature by $d T$ changes the balance between absorption and emission by an amount proportional to a derivative of the Bose-Einstein distribution.

In the second term of the Lindbladian we have distinguished adiabatic (coherent $\delta\rho_\mathrm{coh}$) and non-adiabatic (incoherent $\delta\rho_\mathrm{inc}$) relaxation channels of fermionic density, with the nonadiabatic one depending on a difference between electron and phonon velocities. We make the conjecture that, since changes (due to the phonon environment) in the diagonal elements of the density matrix (a trace of which is proportional to fidelity) are by definition caused by dissipation, only the non-adiabatic processes will cause $\delta \mathcal{F}(T)\neq 0$: If in Eq. (\ref{eq:fidelitydef}) one takes $\rho^B=\rho^A+\delta \rho_\mathrm{inc}$ and then Taylor expand, one finds that the fidelity susceptibility $\chi_{\mathcal{F}}(T)=d\log[\mathcal{F}(T)]/dT \sim d\rho_\mathrm{inc}/dT$.  

Since, in Eq. (\ref{eq:fidelitydef}), we are interested only in the electronic part of the density matrix, $\rho_\mathrm{el}$, we integrate out Eq. (\ref{eq:displacement}) over all possible $q$ (all possible single-phonon emission/absorption events) i.e. $\frac{d\rho_{inc}^{(1)}}{dT}=\sum_{q} \frac{d\rho_\mathrm{inc}}{dn_q}\frac{dn_q}{dT}$ which can be performed analytically:

\begin{widetext}
\begin{multline}\label{eq:athermal-distrib} 
 \frac{d\rho_\mathrm{inc}^{(1)}(\vec{k},T;E_\mathrm{DFT}(T))}{dT} =\\ \frac{v_F(\vec{k},T)}{v_F(\vec{k},T)-c_\mathrm{ph}(T)}\bar{g}_\mathrm{el-ph} f\Big(\frac{E_\mathrm{DFT}(\vec{k},T)}{k_B T}\Big)
 \Bigl\{\Phi\bigl(\exp\{[-E_\mathrm{DFT}(\vec{k},T)-\omega_0]/k_B T\},\alpha,1-2 c_\mathrm{ph}/v_F(\vec{k},T)\bigr)\\-
 \Phi\bigl(\exp\{[-E_\mathrm{DFT}(\vec{k},T)+\omega_0]/k_B T\},\alpha,1+2 c_\mathrm{ph}/v_F(\vec{k},T)\bigr)\Bigr\}.
\end{multline}
\end{widetext}
We require $\alpha>0$ as then the generalized Fermi integral over $q$ turns into an analytic expression involving the Lerch transcendent function, $\Phi(,,)$, which results from integrating out $(1-f_{k-q})$ times the Boltzmann distribution times the power law $q^\alpha$ over a Hankel contour. This formula generalizes past results for Fermi integrals that were expressed as polylogarithms. The advantage of Eq. (\ref{eq:athermal-distrib}) is that it does not constrain any of $c_\mathrm{ph}/V$ or $g_\mathrm{el-ph}/V$ to a small parameter range (nonadiabatic regime) nor does it make any assumptions on thermal distributions [hence it is valid in the intermediate temperature regime where $f(E)$ is certainly not a step function]. Furthermore, the accuracy of this semianalytic approach is not critically dependent on a density of reciprocal space sampling, as it would have been the case in a purely numerical approach. 

\begin{figure}[]
    \centering
    \includegraphics[width=\columnwidth]{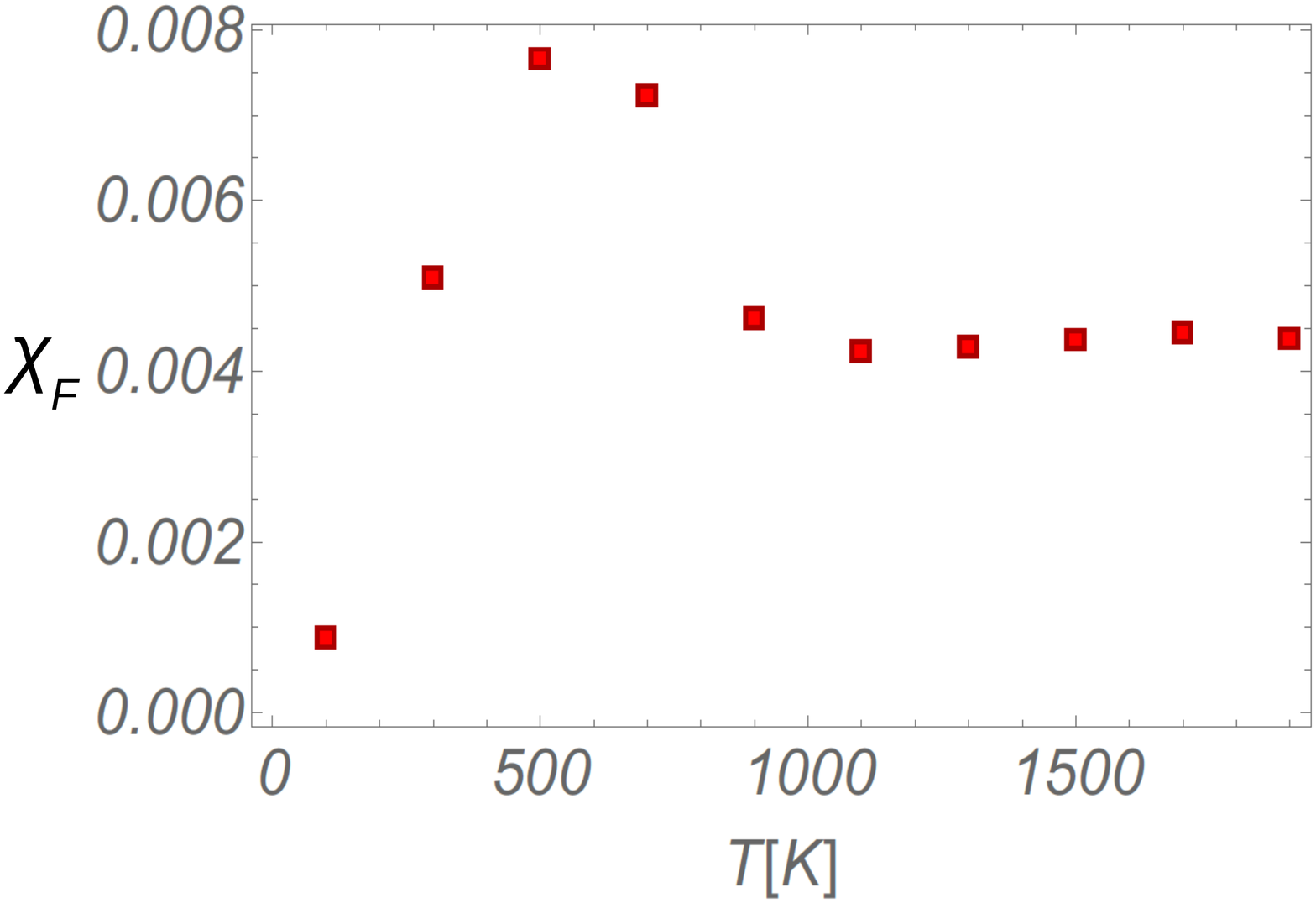}
    \caption{An overall $T$ dependence of quantum fidelity derivative (with a minus sign) obtained by truncating the interactions at $(n=10)$th order and integrating over the entire BZ to find $d \rho_\mathrm{inc}^{(n)}(T)/dT\sim\chi_F(T)$. We see a peak indicating the phase transition.}
    \label{fig:F(T)}
\end{figure}

\begin{figure}[!ht]
    \centering
     \includegraphics[width=\columnwidth]{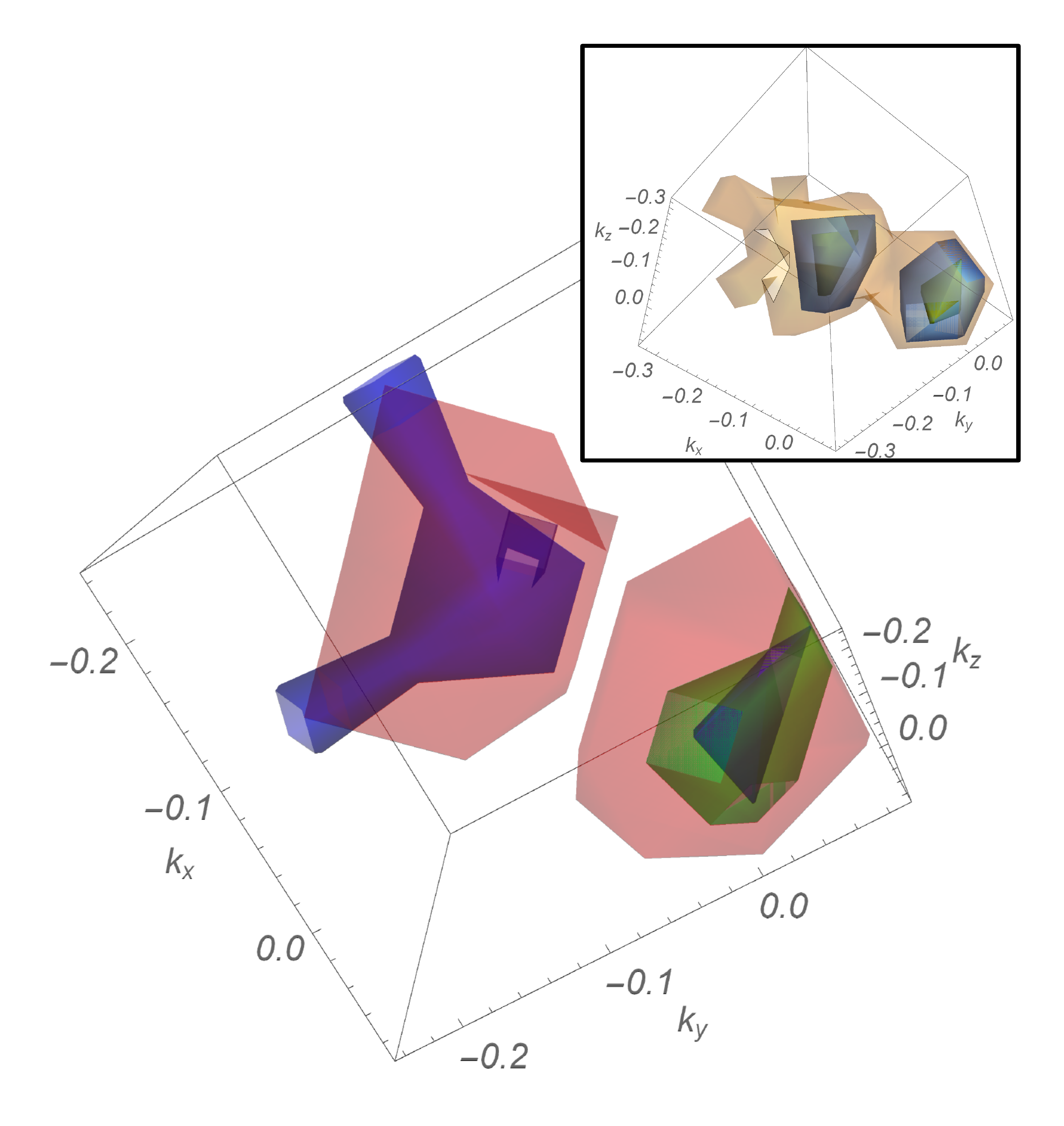}
     \caption{Momentum-resolved fidelity susceptibility $\chi_{\mathcal{F}}(T)$ shown in a region of the BZ around the $L$ point. In the main panel we show isosurfaces corresponding to $\chi_{\mathcal{F}}(T)=0.02$ for $T=300$ K (blue), 500 K (pink) and 900 K (green). In the first two cases there are two minima and for 500 K the isosurface is the largest. The inset shows three isosurfaces $\chi_{\mathcal{F}}(T)=0.3$, $\chi_{\mathcal{F}}(T)=0.03$, $\chi_{\mathcal{F}}(T)=0.01$ for $T=500$ K, showing that the largest values of drag are indeed concentrated close to the minima.}
    \label{fig:drag_fidel}
\end{figure}

The above calculation gives a probability of electronic density shifting into the incoherent part due to an interaction with a \emph{single} phonon. Beyond the weak-coupling regime \cite{Varma-strong-couplCDW} we are interested in a recursive process, where the electronic density shifted by the $n$th interaction is consecutively distorted by the $(n+1)$th interaction. We then need to solve such a recursion problem, i.e., the stationary condition expressed in Eq. (\ref{eq:master}), couples the $|n_\mathrm{ph}\rangle$ state with $|n_\mathrm{ph}\pm 1 \rangle$ states, hence to find the overall stationary state $\hat{L}\{\delta\rho^{(\infty)}\}=0$ we need to find the kernel of such a tridiagonal matrix. The solution $\delta\rho_\mathrm{inc}^{\infty}(k,q;T)$ is a continued fraction. The result from Eq. (\ref{eq:athermal-distrib}), $d\rho_\mathrm{inc}^{(1)}(\vec{k},T;E_\mathrm{DFT}(T))/dT $, can be simply incorporated into the continuous fraction solution for the derivative 
\begin{multline}
   \frac{d\rho_\mathrm{inc}^{(n+1)}(\vec{k},T;E_\mathrm{DFT}(T))}{dT} =\\
    \ContFracOp_{m=1}^n \frac{\bar{g}_\mathrm{el-ph}^m \frac{d\rho_\mathrm{inc}^{(1)}(\vec{k},T;E_\mathrm{DFT}(T))}{dT}}{1+\bar{g}_\mathrm{el-ph}^m\frac{d\rho_\mathrm{inc}^{(1)}(\vec{k}\pm \vec{q_0},T;E_\mathrm{DFT}(\vec{k}\pm \vec{q_0}) \pm m\omega_0)}{dT}},
    \label{eq:contfrac}
\end{multline}
where $\ContFracOp_{m=1}^n$ is a Gauss continued fraction symbol. $\chi_F$ is then calculated by entering the DFT electronic band structure and phonon velocity in Eq. (\ref{eq:athermal-distrib}) and recursively applying Eq. (\ref{eq:contfrac}) until convergence is achieved. The fidelity susceptibility plotted in Fig. \ref{fig:F(T)} peaks between 500 and 700 K, indicating a phase transition of the electron gas, driven by a loss of coherence through the coupling with the TO phonons. The reason for this is that in regions of the Brillouin zone (BZ) with extremely low Fermi velocity the drag effect is the strongest and capable of \emph{nonadiabatically} pulling some carriers away from a coherent single-particle manifold into states strongly coupled with phonons which no longer have a well-defined symmetry property. This mechanism is confirmed by the momentum-resolved $\chi_F(T)$, shown in Fig. \ref{fig:drag_fidel}, where we observe a close connection between $\chi_F(T)$ and the shape of the conduction band: The manifolds of largest drag are either located close to the minimum (minima) or at a large bone-shaped zone at the temperature for which the band curvature changes.

{\it Discussion and conclusions}. Our results show, using quantum fidelity as a measure, that a finite-temperature thermal topological phase transition exists.
This transition, due to loss of coherence of the topological state, is predicted to occur at a temperature lower than the melting one and much lower than the gap-closing temperature as calculated with DFT ($\sim1700$ K). 

The position of the peak in the fidelity susceptibility is robust and remains unaffected by small to moderate changes of all renormalizable parameters such as $g_\mathrm{el-ph}$, $c_\mathrm{ph}$, the exponent $\alpha$, or gap $\Delta(T)$, which implies that it is a generic, stable feature of the system that exist in the regime of intermediate electron-phonon coupling.
Considering band shifts due to additional correlation effects and electron-phonon coupling in the starting single-particle electronic structure might slightly shift the value of the transition temperature, but based on previous works we do not expect this to qualitatively affect the results presented here \cite{Querales-Flores2020}. More specifically, we can distinguish two additional effects. First, the imaginary part of electron-phonon self-energy $\Sigma_\mathrm{el-ph}$ may relax strict kinematic conditions for electron-phonon scattering in Eq. (\ref{eq:athermal-distrib}) and thus increase the drag. We note that in Ref. \cite{Querales-Flores2020} the maximum of $\Sigma_\mathrm{el-ph}(T)$ falls at $T\approx 700$ K, hence this shall enhance the peak in Fig. \ref{fig:F(T)}. Second, there might be an additional contribution to drag due to topological states \cite{Parente-el-ph-topo} present, e.g., on dislocations, an effect that is sample dependent. We note that at the transition these topological states disappear, and thus the additional contributions disappear as well, therefore we expect the signatures of the transition to be even more pronounced.

Our formalism can be generalized to any boson that couples with electrons and violates the symmetry property defining the topological class. This may be low-energy interband plasmons for systems defined by parity symmetry, or paramagnons for systems with strong Rashba interactions.  
The only conditions that the boson must fulfill are the $g_\mathrm{el-ph}\sim q^{\alpha}$ with $\alpha>0$ and its dispersive character $\omega(q)=c_\mathrm{ph} q$ for a range of $q$. While these are typical for acoustic phonons, they are also obeyed by symmetry-breaking TO phonons in some incipient ferroelectrics. 
While the position of the phase transition is determined by the change of band curvature, the amplitude of the associated peak is related to the TO phonon's $c_\mathrm{ph}$ and $g_\mathrm{el-ph}$. Overall, the topological thermal phase transition should apply to a much broader class of materials than only the topological crystal insulators studied here.

PAP would like to acknowledge funding from the Diputaci\'on Foral de Gipuzkoa through Grant No. 2020-FELL-000005-01 and the Spanish Ministry for Science and Innovation through Grant No. PID2019-107338RB-C61. PC would like to acknowledge funding from the MSCA No. 847639 and EPSRC EP/V02986X.

\appendix
\section{Technical details of the Density Functional theory calculations}\label{sec:AppDFT}
\label{sec:app_DFT}
DFT calculation were carried out with the code {\sc siesta} \cite{soler2002} using the generalized gradient approximation (GGA) with the Perdew-Burke-Ernzerhof (PBE) \cite{Perdew1996} parametrization of the exchange and correlation functional. We have used fully nonlocal two-projector norm-conserving pseudopotentials from the {\sc PseudoDojo} data base \cite{VanSetten2018} generated with the ONCVPSP code \cite{Hamann2013}. For the basis functions we utilized a set of numerical atomic orbitals optimized following the prescriptions of Ref. \cite{Junquera2001}, with single-$\zeta$ semicore $d$ states and double-$\zeta$ valence states. The electronic density, Hartree, and exchange-correlation potentials are computed in a uniform real space grid, with an equivalent plane-wave cutoff of 800 Ry; and Brillouin zone sampling was carried out with a $12\times 12\times 12$ Monkhorst-Pack mesh \cite{Monkhorst1976} for the self-consistent calculations.

One caveat about our analysis based on DFT density matrices is that in any DFT calculation a change in the lattice parameter changes the basis set in which the Kohn-Sham states are expanded, whether due to the shift of the functions in real space in the case of atomic orbitals, or due to the change in the reciprocal lattice vectors in the case of plane waves. Here, we choose to neglect the drift in the fidelity due to changes in the basis set under the assumption that for sufficiently small changes in the lattice parameter, the off-diagonal elements of the transformation matrix become negligible. The negative result in Fig. 2 in the main document seems to support this assumption.

Estimating the effect of additional electron correlation and band shifts due to electron-phonon interaction on the topological phase transition studied here is beyond the scope of this work. Nevertheless, the effect of these contributions to the gap-closing temperature gives us an idea of the magnitude of the sifts in temperature values that we may expect. Using the results presented in Ref. \cite{Querales-Flores2020} we can obtained the extrapolated values for the gap-closing temperature shown in Tab. \ref{tab:transitionT}. The estimated gathered in Tab. \ref{tab:transitionT} show that additional electronic correlation and electron-phonon coupling have contribute to shift the gap-closing temperature in opposite directions, partially canceling each other. In any case, this value remains significantly larger than the melting temperature of SnTe. This test justifies the use of DFT and the PBE exchange of correlation functional for this study.

\begin{table}[]
  \begin{center}
    \begin{tabular}{lccc}
    \hline \hline
     & PBE & $G_0W_0$ & $G_0W_0$ + e-ph \\
    \hline
     $T_{\Delta=0}$ (K) & 1733 & 2020  & 1515  \\
    \hline \hline
    \end{tabular}
    \caption{Temperatures at which the SnTe band gap closes, as predicted with the PBE functional, $G_0W_0$ approximation and $G_0W_0$ including the band renormalization due to electron-phonon coupling. Values obtained from the extrapolation of data presented in [Phys. Rev B 101, 235206 (2020)]. }
    \label{tab:transitionT}
  \end{center}
\end{table}

\begin{figure*}
    \centering
    \includegraphics[width=\textwidth]{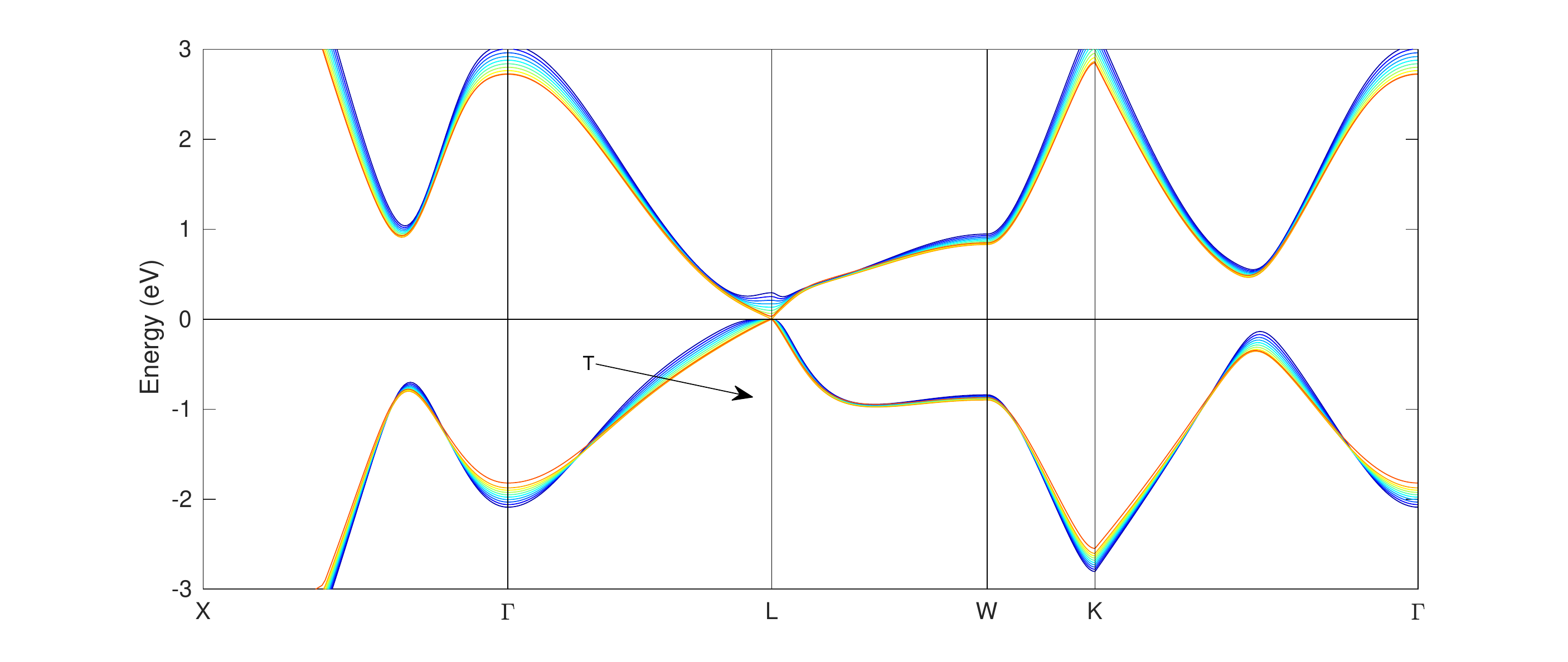}
    \caption{Evolution of the band structure of SnTe with temperature, as calculated with DFT. The temperature, $T$ varies from 100 to 1900 K in 200-K steps.}
    \label{fig:img1}
\end{figure*}

\section{Technical details of drag propensity calculations}
\label{sec:app_drag}
We aim to compute the propensity of the electronic density to shift into incoherent part $\Delta \rho_\mathrm{inc}$ as the temperature increases. Our input are eigen-energies and electronic dispersion $\vec{v}$ (momentum derivatives of eigen-energies) at each point of 3D Brillouin zone. We first consider a single event of phonon's emission-absorption for each electronic $k$-state, i.e. we apply Eq. (\ref{eq:athermal-distrib}) to each $k$-state. We made only two physical assumption in our reasoning. The first assumption for the solution of the problem given by Eq. (\ref{eq:hamilt}) is that we assume that for bosons the \emph{change} of level occupancy is given by Boltzmann thermal factor, this is valid provided thermalization hypothesis of bosonic sub-system holds (any effective temperature can be accommodated in our formalism); while for fermions we assume that the process requires participation of an entire fermion, so even if $\hat{\rho_\mathrm{el}}$ is e.g. one for fractionalized Fermi liquid, the emission-absorption process requires re-fermionization. We do not set any other constrains for the nature of the ground state. The second assumption is by invoking vertex correction (Fresnel wave diffraction) we assume a strong preference for small angle electron-phonon scattering, $\vec{k}||\vec{k}+\vec{q}$. Thus, in Eq. (\ref{eq:athermal-distrib}) we take a projection of the velocity $v_\mathrm{scatt}=\vec{v}\cdot \vec{k}$. Since we take isotropic $c_\mathrm{ph}$ and $g_\mathrm{el-ph}$ the rest of the implementation is straightforward. Please note that entire influence of higher order electron-phonon coupling, in a ladder approximation, can be accommodated through renormalization of $\bar{g}_\mathrm{el-ph}$ and $\alpha$ hence it can be in principle accounted for in Eq. (\ref{eq:contfrac}).

We then move to the problem of quantum trajectories with multiple phonon emission-absorption events. To this end we extend the electronic density matrix by adding another dimension that describes the number of phonon associated with electron's motion (these would be an analogue of Floquet bands in the pumping problem). This is a pseudo-spin degree of freedom attached to each fermion, with the parity of the pseudo spin that determine the symmetry with respect to the crystal mirror plane. The off-diagonal elements are described by Eq. (\ref{eq:athermal-distrib}) (where we take an average absorption-emission from neighboring $k$-states along the $\vec{k}$ ray). We need to diagonalize this tridiagonal matrix, and then the lowest eigenstate will give us the desired $\partial_T \rho_\mathrm{inc}^{(n)}(\vec{k};T)$. The diagonalization of tridiagonal matrix, through a ratio of continuant polynomials, leads to the continuous fraction expression given in Eq. (\ref{eq:contfrac}).

It should be emphasized that the origin of the observed maximum is not due to any of the prefactors in Eq. (\ref{eq:athermal-distrib}) but it is a generic property of the Lerch transcendent function $\Phi$ in the vicinity of $v_F(\vec{k},T)\rightarrow 2c_\mathrm{ph}$. The inspection of the recursive re-summation is then necessary only to check whether this maximum survives in the higher order perturbation theory or decays. Our results indicated that it is even enhanced. It should be however mentioned that in these higher order processes we took an assumption of Fermi-Dirac distribution of particles in the polaronic bands, with unrenormalized $V_F$ and $g_\mathrm{el-ph}$. Our formalism may account for these generalization, which we postpone to further studies.

\bibliographystyle{apsrev4-1} 
\bibliography{library_pablo.bib,piotr-refs.bib}

\end{document}